\documentstyle[preprint,array,multicol,eqsecnum,prd,aps,graphicx]{revtex}

\begin{document}

\preprint{US-03-04, STUPP-03-173}

\title{$R$-Parity Violation in a SUSY GUT Model\\
and Radiative Neutrino Masses}

\author{\bf Yoshio Koide\thanks{
E-mail address: koide@u-shizuoka-ken.ac.jp}
and Joe Sato\thanks{E-mail address: joe@phy.saitama-u.ac.jp
}$^{(a)}$}

\address{
Department of Physics, University of Shizuoka, 
52-1 Yada, Shizuoka 422-8526, Japan \\
(a) Department of Physics, Faculty of Science, Saitama University,  
Saitama, 338-8570, Japan}


\maketitle
\begin{abstract}
Within the framework of an SU(5) SUSY GUT model, a mechanism
which effectively induces $R$-parity-violating terms below the 
unification energy scale $M_X$ is proposed.
The model has matter fields $\overline{5}_{L(+)}+10_{L(-)}$
and Higgs fields $H_{(-)}$ and $\overline{H}_{(+)}$ in addition
to the ordinary Higgs fields $H_{(+)}$ and $\overline{H}_{(-)}$ 
which contribute to the Yukawa interactions, where $(\pm)$ denote 
the transformation properties under a discrete symmetry Z$_2$.
The Z$_2$ symmetry is only broken by the $\mu$-term 
$\overline{H}_{(+)}H_{(-)}$ softly, so that the 
$\overline{5}_{(+)}\leftrightarrow \overline{H}_{(+)}
\leftrightarrow \overline{H}_{(-)}$ mixing appears at
$\mu < m_{SB}$, and $R$-parity violating terms 
$\overline{5}_L\overline{5}_L 10_L$ are effectively induced
from the Yukawa interactions $\overline{H}_{(-)}\overline{5}_{L(+)}
10_{L(-)}$, i.e. the effective coupling constants $\lambda_{ijk}$
of $\nu_{Li} e_{Lj} e_{Rk}^c$ and $\nu_{Li} d_{Rj}^c d_{Lk}$
are proportional to the mass matrices $(M_e^*)_{jk}$ and
$(M_d^\dagger)_{jk}$, respectively.  The parameter regions 
which are harmless for the proton decay are investigated.
Possible forms of the radiatively induced neutrino mass matrix 
are also investigated.
\end{abstract}


\maketitle



\section{introduction}
\label{Sec1}

The origin of the neutrino mass generation is still a mysterious
problem in the unified understanding of the quarks and leptons.
As the origin, from the standpoint of a grand unification theory 
(GUT), currently, the idea of the so-call seesaw 
mechanism\cite{seesaw} is influential.
On the other hand, an alternative idea that the neutrino masses 
are radiatively induced is still attractive. 
As an example of such a model, the Zee model \cite{Zee} is well 
known.
Regrettably,  the original Zee model is not on the 
framework of GUT. 
A possible idea   to embed the Zee model 
into GUTs is to identify the Zee scalar  $h^+$ as the slepton 
$ \widetilde{e}_R$ in an $R$-parity-violating supersymmetric 
(SUSY) model \cite{R_SUSY}. 
However, usually, it is accepted that SUSY models with $R$-parity 
violation are incompatible with a GUT scenario, 
because the $R$-parity-violating interactions induce 
proton decay \cite{Smirnov95,Smirnov96}.
By the way, there is another problem in a GUT scenario, 
i.e. how to give doublet-triplet splitting in SU(5) 5-plet
Higgs fields.
There are many ideas to solve  this problem \cite{DTsplitting}.
Although these mechanisms are very attractive,
in the present paper, we will take another 
choice, that is, fine tuning of parameters:
we consider a possibility that a mechanism which provides 
the doublet-triplet splitting gives a suppression of 
the $R$-parity violating terms with baryon number violation  
while it gives visible contributions of the doublet component 
to the low energy phenomena (neutrino masses, lepton flavor 
processes, and so on)\cite{Smirnov95}.
In the present paper, we will try 
to give an example of such a scenario.

In the present paper, in order to suppress the proton decay, 
a discrete symmetry Z$_2$ is introduced.
The essential idea is as follows:
we consider  matter fields $\overline{5}_{L(+)} +10_{L((-)}$ of
SU(5) and two types of SU(5) $5$-plet and $\overline{5}$-plet Higgs 
fields $H_{(\pm)}$ and $\overline{H}_{(\pm)}$, 
where $(\pm)$ denote the transformation 
properties under a discrete symmetry Z$_2$ (we will call it 
``Z$_2$-parity" hereafter).
The superpotential in the present model is given by
\begin{equation}
W=W_Y + W_H + W_{mix} \ ,
\end{equation}
where $W_Y$ denotes  Yukawa interactions
\begin{equation}
W_Y  =  \sum_{i,j} (Y_u)_{ij} H_{(+)} 10_{L(-)i} 10_{L(-)j} 
     +  \sum_{i,j} (Y_d)_{ij} \overline{H}_{(-)}
\overline{5}_{L(+)i}10_{L(-)j} \ .
\end{equation}
Under the discrete symmetry Z$_2$, $R$-parity violating
terms $\overline{5}_{L(+)} \overline{5}_{L(+)} 10_{(-)}$
are exactly forbidden.
The discrete symmetry Z$_2$ is softly violated only by
the following $\mu$-terms 
\begin{equation}
W_H  =   \overline{H}_{(+)} (m_+ +g_+ \Phi) {H}_{(+)} 
  + \overline{H}_{(-)} (m_- +g_- \Phi){H}_{(-)} 
  +   m_{SB}  \overline{H}_{(+)} H_{(-)}\ ,
\end{equation}
where $\Phi$ is an SU(5) 24-plet Higgs field with the
vacuum expectation value (VEV) $\langle\Phi\rangle = v_{24}
{\rm diag}(2,2,2,-3,-3)$, and it has been introduced in order to 
give doublet--triplet splittings in the SU(5) 5- and 
$\overline{5}$-plets Higgs fields at an energy scale $\mu < M_X$
($M_X$ is an SU(5) unification scale).
The Z$_2$-parity is violated only by\footnote{
The Z$_2$ symmetry can softly violated not only by the
term $\overline{H}_{(+)} H_{(-)}$, but also by terms
$\overline{H}_{(-)} H_{(+)}$ and $\overline{5}_{L(+)1} H_{(-)}$.
However, in the present scenario, the existence of 
$\overline{H}_{(+)} H_{(-)}$ is essential.
The details are discussed Appendix A.
}
 the term $\overline{H}_{(+)} H_{(-)}$. 
Note that $ H_{(-)}$ and $\overline{H}_{(+)}$ in the $m_{SB}$-term 
do not contribute to the Yukawa interaction (1.2) directly,
so that proton decay via the dimension-5 operator is suppressed
in the limit of $m_{SB} \rightarrow 0$.
(A similar idea, but without Z$_2$ symmetry, has been
proposed by Babu and Barr \cite{twoHiggs}.)
The terms $W_{mix}$ have been introduced in order 
to bring the $\overline{H}_{(+)} \leftrightarrow \overline{5}_{(+)}$ 
mixing:
\begin{equation}
W_{mix} =  \sum_i \overline{5}_{L(+)i} \left( b_i m_5 
+ c_i g_5 \Phi \right) H_{(+)} \ ,
\end{equation}
where $\sum_i |b_i|^2=\sum_i |c_i|^2=1$.
At the energy scale $\mu  <   M_X$, the terms $W_H + W_{mix}$ are effectively
given by
\begin{eqnarray}
W_H + W_{mix} &=& \sum_{a=2,3} m^{(a)}_+ 
\left[\overline{H}^{(a)}_{(+)} \cos \alpha^{(a)} + \sum_i d_i  
\overline{5}^{(a)}_{L(+)i} \sin \alpha^{(a)}
\right] H^{(a)}_{(+)} \nonumber \\
&&+ \sum_{a=2,3} m^{(a)}_{-}  \overline{H}^{(a)}_{(-)} H^{(a)}_{(-)}
+ m_{SB} \sum_{a=2,3} \overline{H}_{(+)}^{(a)}  H^{(a)}_{(-)} \ ,
\end{eqnarray}
where $ \sum_{i} |d_i|^2 =1$ , the index (a) denotes that the fields with 
(2) and
(3) are doublet and triplet components of 
SU(5)$\rightarrow$SU(2)$\times$SU(3), 
respectively, and
\begin{equation}
m^{(2)}_{+} \cos \alpha^{(2)} = m_{+} -3g_+ v_{24} \ , \ \ 
m^{(3)}_{+} \cos \alpha^{(3)} = m_{+} +2g_+ v_{24} \ ,
\end{equation}
\begin{equation}
m^{(2)}_{+} \sin \alpha^{(2)} d_i = m_{5}b_i -3g_5 v_{24} c_i \ , \ \ 
m^{(3)}_{+} \sin \alpha^{(3)} d_i = m_5 b_i + 2g_5 v_{24} c_i \ ,
\end{equation}
\begin{equation}
m^{(2)}_{-}  = m_{-} -3g_{-}  v_{24} \ , \ \ 
m^{(3)}_{-}  = m_{-} +2g_{-} v_{24} \ .
\end{equation}
Therefore, the $m_{SB}$-term together with $m_+ \sin\alpha$-term
induces the $\overline{H}_{(-)}\leftrightarrow\overline{5}_{L(+)}$
mixing, so that the $R$-parity violating terms 
 $\overline{5}_{L}\overline{5}_{L}10_{L}$
are generated from the Yukawa interactions
$\overline{H}_{(-)}\overline{5}_{L(+)}10_{L(-)}$.
The coupling constants $\lambda_{ijk}$ of $\overline{5}_i
\overline{5}_j 10_k$ will be proportional to the charged lepton
mass matrix $(M_e^*)_{jk}$ or down-quark mass matrix
$(M_d^\dagger)_{jk}$. (The details are discussed in the next
section II.)
As we demonstrate in Sec.~II, we can show that 
the mixing $\overline{5}_{L(+)} \leftrightarrow \overline{H}_{(-)}$
is negligibly small for the colored sector, while it is
sizable for SU(2)-doublet sector.

The parameters in the present model need fine-tuning.
For example, we will find that a large value of $m_{SB}$ is not 
acceptable, because for such a large value of $m_{SB}$ the proton 
decay due to the dimension five operator becomes visible.
On the other hand, we will find  that a smaller value of $m_{SB}$
leads  to a small  bottom quark mass, so  that a small value
of $m_{SB}$ is not acceptable.
We will take $m_{SB}\sim 10^{14}$ GeV.
In Sec.~III, we will investigate the parameter regions 
which are harmless for the proton decay.
In Sec.~IV, we will investigate a possible form of 
the radiatively induced neutrino mass matrix due  to  
the $R$-parity violation term
$\overline{5}_{L}\overline{5}_{L}10_{L}$.
The radiatively induced neutrino mass matrix $M_\nu^{rad}$
will be expressed by the sum of two rank-1 matrices.
On the other hand, we also have contributions $M_{\tilde{\nu}}$
from VEVs $\langle \tilde{\nu}_i\rangle$ of the sneutrinos
to the neutrino mass matrix $M_\nu$.
In Sec.~V, a possible form of $M_\nu=M_{\tilde{\nu}}+
M_\nu^{rad}$ is discussed from the phenomenological point
of view.
Finally, Sec.~VI will be devoted to the summary.

\section{$\overline{H}_{(-)}$-$\overline{5}_{(+)}$
mixing} 
\label{sec:2}

In order to suppress the proton decay, we want to take 
$m^{(2)}_{+} \sim M_W$ with a sizable $\alpha^{(2)}$,
but $m^{(3)}_{+} \sim M_X$ with a negligibly small $\alpha^{(3)}$.
However, from the relations (1.6) and (1.7), 
we obtain the relation
\begin{equation}
d_i \tan\alpha^{(3)} = \frac{m_5 b_i +2 g_5 v_{24} c_i}{
m_+ +2g_+ v_{24} } =
\frac{m_+^{(2)} \sin\alpha^{(2)}d_i +5 g_5 v_{24} c_i}{
m_+^{(2)} \cos\alpha^{(2)} +5 g_+ v_{24} } \ .
\end{equation}
The requirement $|\alpha^{(3)}| \alt M_W/M_X$ leads to the 
constraint $|g_5| \alt M_W/M_X$ for $|g_+|\sim 1$.
We do not like to introduce such a small dimensionless 
parameter $g_5$.
Therefore, for simplicity, we will put $g_5=0$ hereafter.
Then, without loss of generality, we can put
\begin{equation}
\overline{5}'_{L(+)1} = \sum_i b_i \overline{5}_{L(+)i}
\end{equation}
where $\overline{5}'_{L(+)1}$ does not mean the observed first 
generation particle. 
(Hereafter, for convenience, we denote $\overline{5}'_{L(+)1}$
as $\overline{5}_{L(+)1}$ simply.
The effective parameters $m^{(a)}_{+}$, $m^{(a)}_{-}$ and 
$\alpha^{(a)}$ are given as follows:
\begin{eqnarray}
&m^{(2)}_{+} = \sqrt{(m_{+} -3g_+v_{24})^2 + m^2_5} \ ,  \ \ \ \ 
& m^{(3)}_{+} = \sqrt{(m_{+} +2g_+v_{24})^2 + m^2_5} \ ,
\nonumber \\
&m^{(2)}_{-} = m_{-} -3g_{-} v_{24} \ ,  \ \ \ \ 
& m^{(3)}_{-} = m_{-} +2g_{-} v_{24} \ , \\
&\tan \alpha^{(2)} =\frac{m_5}{m_{+} -3g_+v_{24}} \simeq 
\frac{m_5}{m_+^{(2)}}\ ,  
& \tan \alpha^{(3)} =\frac{m_5}{m_{+} +2g_+v_{24}} \simeq 
\frac{m_5}{m_+^{(3)}} \ . \nonumber 
\end{eqnarray}
We will take
\begin{eqnarray}
& m^{(2)}_{+} \sim M_W \ , \ \ \ \ \ \
m^{(3)}_{+} \sim M_X \ ,
\nonumber \\
& m^{(2)}_{-} \sim M_{I} \ , \ \ \ \ \ 
m^{(3)}_{-} \sim M_X \ , \\
& \tan \alpha^{(2)} \sim \frac{m_5}{M_W} \ , \ \ \ \ \ 
\tan \alpha^{(3)} \sim \frac{m_5}{M_X} \ ,\nonumber
\end{eqnarray}
where $M_I \sim 10^{14}$ GeV and $m_5 \sim 10^{1}$ GeV as we state later.
The mass matrix in the basis of 
$(\overline{H}_{(-)}, \overline{H}_{(+)},\overline{5}_{L(+)1})$ 
and $(H_{(+)},H_{(-)})$
is given by
\begin{equation}
M = \left(
\begin{array}{cc}
0 & m_- \\
m_+ \cos\alpha & m_{SB} \\
 m_+ \sin\alpha & 0  
\end{array} \right) \ .
\end{equation}
Here and hereafter, for simplicity, we drop the index $(a)$.
The mass matrix (2.5) is diagonalized as
\begin{equation}
\overline{U}^\dagger M {U} = D \equiv \left(
\begin{array}{cc}
m_1 & 0  \\
0 & m_2 \\ 
0 & 0
\end{array} \right) \ ,
\end{equation}
where $U$ and $\overline{U}$ are unitary operators which diagonalize
$M^\dagger M$ and $M M^\dagger$ as
\begin{equation}
U^\dagger  M^\dagger M U = \left(
\begin{array}{cc}
m_1^2 & 0 \\
0 & m_2^2  
\end{array} \right) \ ,
\end{equation}
and
\begin{equation}
\overline{U}^\dagger M M^\dagger \overline{U} = \left(
\begin{array}{ccc}
m_1^2 & 0 & 0 \\
0 & m_2^2 & 0 \\
0 & 0 & 0
\end{array} \right) \ ,
\end{equation}
respectively.
Note that the  matter field $\overline{5}_{L1}^{\prime}$
is still massless, and also note that it is not in the eigenstate 
of the Z$_2$ parity.

The mixing matrix $U$ is easily obtained from the diagonalization of
\begin{equation}
M^\dagger M = \left(
\begin{array}{cc}
|m_+|^2 & m_{SB} m_+^* \cos\alpha \\
m_{SB}^* m_+ \cos\alpha & |m_{SB}|^2 +|m_-|^2 
\end{array} \right) \ .
\end{equation}
For real $m_1$, $m_{SB}$ and $m_\pm$, we obtain
\begin{equation}
U = \left(
\begin{array}{cc}
\cos\theta_u & \sin\theta_u \\
-\sin\theta_u & \cos\theta_u
\end{array} \right) \ ,
\end{equation}
\begin{equation}
\tan 2\theta_u =  
\frac{2 m_{SB} m_+\cos\alpha}{m_{SB}^2 +m_-^2 -m_+^2} \ ,
\end{equation}
\begin{equation}
m_1^2 = \frac{1}{2}\left( m_{SB}^2+m_+^2+m_-^2 \right)
-\frac{1}{2} Q \ ,
\end{equation}
\begin{equation}
m_2^2 = \frac{1}{2}\left( m_{SB}^2+m_+^2+m_-^2 \right)
+\frac{1}{2} Q \ ,
\end{equation}
where 
\begin{equation}
Q=(m_{SB}^2 -m_+^2+m_-^2)\cos 2\theta_u +
2m_{SB} m_+ \cos\alpha \sin 2\theta_u \ .
\end{equation}
When we define
\begin{equation}
A\equiv m_{SB}^2 -m_+^2+m_-^2 \ , \ \ \ 
B\equiv 2 m_{SB} m_+ \cos\alpha \ ,
\end{equation}
\begin{equation}
\cos 2\theta_u = \frac{A}{\sqrt{A^2+B^2}} \ , \ \ \ 
\sin 2\theta_u = \frac{B}{\sqrt{A^2+B^2}} \ ,
\end{equation}
the quantity $Q$ is given by 
\begin{equation}
Q=\sqrt{A^2+B^2}=
\sqrt{
\left[ m_{SB}^2 +(m_+-m_-)^2\right]
\left[ m_{SB}^2 +(m_++m_-)^2\right]
-4 m_{SB}^2 m_+^2 \sin^2 \alpha } \ .
\end{equation}

The rotation $\overline{U}$ is also obtained from the diagonalization of
\begin{equation}
M M^\dagger  = \left(
\begin{array}{ccc}
m_-^2 & m_{SB} m_-  & 0 \\
m_{SB} m_-  & m_{SB}^2 + m_+^2 \cos^2\alpha & 
m_+^2 \cos\alpha \sin\alpha \\
0  & m_+^2 \cos\alpha \sin\alpha  & m_+^2 \sin^2\alpha 
\end{array} \right) \ .
\end{equation}
The mixing matrix elements $\overline{U}_{i3}$ are easily obtain
as follows:
\begin{equation}
\overline{U}_{13}=  \frac{1}{N_3}m_{SB} \sin\alpha \ ,
\end{equation}
\begin{equation}
\overline{U}_{23}= - \frac{1}{N_3} m_- \sin\alpha \ ,
\end{equation}
\begin{equation}
\overline{U}_{33}=  \frac{1}{N_3} m_- \cos\alpha \ ,
\end{equation}
where
\begin{equation}
N_3^2 =- m_-^2 + m_{SB}^2 \sin^2 \alpha \ .
\end{equation}

Other matrix elements are obtained as follows:
We express the mixing matrix $\overline{U}$ as
\begin{equation}
\overline{U} = \left( 
\begin{array}{ccc}
c_{13} c_{12} & c_{13} s_{12} & s_{13} \\
-c_{23} s_{12} -s_{23} c_{12} s_{13} &
c_{23} c_{12} -s_{23} s_{12} s_{13} & s_{23} c_{13} \\
s_{23} s_{12} -c_{23} c_{12} s_{13} &
-s_{23} c_{12} -c_{23} s_{12} s_{13} & c_{23} c_{13} 
\end{array} \right) \ ,
\end{equation}
where $s_{ij}=\sin\theta_{ij}$ and $c_{ij}=\cos\theta_{ij}$.
Then, by comparing (2.23) with (2.19)--(2.21), we obtain
\begin{equation}
s_{13} = \overline{U}_{13}= \frac{m_{SB} \sin\alpha}{\sqrt{
m_-^2 + m_{SB}^2 \sin^2\alpha}} \ , \ \ 
c_{13}= \frac{1}{ \sqrt{ 1+(m_{SB}/m_-)^2\sin^2\alpha}} \ ,
\end{equation}
\begin{equation}
s_{23} =\frac{\overline{U}_{23}}{c_{13}} =- \sin\alpha \ , \ \ 
c_{23}= \cos\alpha \ .
\end{equation}
By using the relation
$(M^\dagger M)_{11}=\overline{U}_{11}\overline{U}_{11} (m'_1)^2
+ \overline{U}_{12} \overline{U}_{12} (m'_2)^2$, 
the mixing angle $\theta_{12}$
is obtained as follows:
\begin{eqnarray}
\cos2\theta_{12}& =& \frac{1}{m_2^2-m_1^2 }
\left[ m_1^2+m_2^2 -2 \frac{m_-^2}{c_{13}^2} \right]
 =\frac{1}{Q}
( m_{SB}^2 +m_+^2 -m_-^2 - 2 m_{SB}^2 \sin^2 \alpha ) 
\nonumber \\
& =& - \frac{m_-^2 -m_2^2\cos 2\alpha - m_+^2}{
 \sqrt{ m_-^4+ 2 (m_2^2-m_+^2)m_-^2+m_2^4 
+ 2 m_2^2 m_+^2 \cos 2\alpha +m_+^4 } } \ . 
\end{eqnarray}
Note that $\cos 2\theta_{12} \simeq -1$ for $m_-^2\gg m_{SB}^2, 
m_+^2$, so that $\theta_{12} \simeq \pi/2$.

Since the physical fields $(\overline{H}_{1}, \overline{H}_{2}, 
\overline{5}_{L1}^{\prime},\overline{5}_{L2}^{\prime},
\overline{5}_{L3}^{\prime})$ 
are given by
\begin{equation}
\left(
\begin{array}{c}
\overline{H}_{(-)} \\
\overline{H}_{(+)} \\
\overline{5}_{L(+)1} \\
\overline{5}_{L(+)2} \\
\overline{5}_{L(+)3} \\
\end{array} \right) =  \left(
\begin{array}{ccccc}
\overline{U}_{11} & \overline{U}_{12} & \overline{U}_{13} & 0 & 0 \\
\overline{U}_{21} & \overline{U}_{22} & \overline{U}_{23} & 0 & 0 \\
\overline{U}_{31} & \overline{U}_{32} & \overline{U}_{33} &  0  & 0  \\
0 & 0 & 0 & 1 & 0  \\
0 & 0 & 0 & 0 & 1 \\
\end{array}\right) \left(
\begin{array}{c}
\overline{H}_{1} \\
\overline{H}_{2} \\
\overline{5}_{L1}^{\prime} \\
\overline{5}_{L2}^{\prime} \\
\overline{5}_{L3}^{\prime} \\
\end{array} \right)  \ ,
\end{equation}
the Yukawa interactions 
$\overline{H}_{(-)}\overline{5}_{L(+)}10_{L(-)}$
lead to the effective Yukawa interactions at a low
energy scale
\begin{equation}
(Y_d)_{ij} \overline{H}_{1} \left[ \delta_{i1}
(\overline{U}_{11} \overline{U}_{33} 
-\overline{U}_{13} \overline{U}_{31})\overline{5}_{L1}^{\prime} +
\overline{U}_{11}(\delta_{i2}\overline{5}_{L2}^{\prime} 
+\delta_{i3} \overline{5}_{L3}^{\prime} )   
 \right] 10_{L(-)j} \ ,
\end{equation}
and
the induced $R$-parity violating terms 
\begin{equation}
(Y_d)_{ij} \overline{U}_{13} \overline{5}_{L1}^{\prime} 
\left( 
\delta_{i1} \overline{U}_{33} \overline{5}_{L1}^{\prime}+
\delta_{i2}\overline{5}'_{L2} +\delta_{i3}\overline{5}'_{L3} 
\right) 
10_{L(-)j}  \ ,
\end{equation}
where we have assumed that $|m_1|<<|m_2|$, i.e. the Higgs field 
surviving at a low energy scale
is not $\overline{H}_{2}$, but $\overline{H}_{1}$.

The effective Yukawa interactions (2.28) give the fermion mass 
matrices
\begin{equation}
(M_e^*)_{ij}= \left\{
\begin{array}{ll}
( \overline{U}_{11}^{(2)} \overline{U}_{33}^{(2)} 
-\overline{U}_{13}^{(2)} \overline{U}_{31}^{(2)}) (Y_d)_{ij} v_d 
=\overline{U}_{22}^{(2)*}  (Y_d)_{ij} v_d & {\rm for}\ i=1 \ , \\
\overline{U}_{11}^{(2)}  (Y_d)_{ij} v_d  & {\rm for}\ i=2,3 \ , \\
\end{array} \right.
\end{equation}
\begin{equation}
(M_d^\dagger)_{ij}= \left\{
\begin{array}{ll}
( \overline{U}_{11}^{(2)} \overline{U}_{33}^{(3)} -
\overline{U}_{13}^{(3)} \overline{U}_{31}^{(2)}) (Y_d)_{ij} 
v_d & {\rm for}\ i=1 \ , \\
\overline{U}_{11}^{(2)} (Y_d)_{ij} v_d
& {\rm for}\ i=2,3 \ , \\
\end{array} \right.
\end{equation}
where $v_d= \langle \overline{H}_1\rangle$, and, 
in (2.30), we have used the general formula 
$U_{ik} U_{jl} - U_{il} U_{jk} = \varepsilon_{ijm} \varepsilon_{kln}
U_{mn}^*$ for an arbitrary $3\times 3$ unitary matrix $U$.
Note that in the present model, the relation 
$M_d=M_e^T$ does not hold.

{}From the $R$-parity violating terms (2.29), we obtain 
coefficients $\lambda_{ijk}^{(2,2)}$, $\lambda_{ijk}^{(2,3)}$,
$\lambda_{ijk}^{(3,2)}$ and $\lambda_{ijk}^{(3,3)}$, which are
the coefficients of the interactions
$(\nu_{L1}e_{Li} -e_{L1} \nu_{Li}) e_{Rj}^c$,
$(\nu_{L1} d_{Ri}^c d_{Lj} - e_{L1} d_{Ri}^c u_{Lj})$,
$(d_{R1}^c e_{Li} u_{Lj}-d_{R1}^c \nu_{Li} d_{Lj})$, and
$\varepsilon_{\alpha\beta\gamma} d_{R1}^{c\alpha}
d_{Ri}^{c\beta} u_{Rj}^{c\gamma}$, respectively, as follows:
\begin{eqnarray}
\lambda_{11j}^{(2,2)} &=& 0 \ , 
\ \ \ \ \ \ \ \ \ \ \ \ \ \ 
\lambda_{1ij}^{(2,2)} = \kappa 
(M_e^*)_{ij}/v_d \ \ \ 
(i=2,3) , \\
\lambda_{11j}^{(2,3)} &=&
\frac{\kappa (M_d^\dagger)_{1j}/v_d}{
1-\xi\kappa \overline{U}_{31}^{(2)}/\overline{U}_{33}^{(3)}} \ , 
\ \ 
\lambda_{1ij}^{(2,3)} = \kappa
(M_d^\dagger)_{ij}/v_d \ \ \ 
(i=2,3) , \\
\lambda_{11j}^{(3,2)} &=& 
\frac{\xi \kappa (M_e^*)_{1j}/v_d}{
1-\kappa \overline{U}_{31}^{(2)}/\overline{U}_{33}^{(2)}} \ , 
\ \ 
\lambda_{1ij}^{(3,2)} = 
\xi \kappa (M_e^*)_{ij}/v_d \ \ \ 
 (i=2,3) , \\
\lambda_{11j}^{(3,3)} &=& 0 \ , 
\ \ \ \ \ \ \ \ \ \ \ \ \ \ 
\lambda_{1ij}^{(3,3)} =
\xi \kappa (M_d^\dagger)_{ij}/v_d \ \ \
(i=2,3) , 
\end{eqnarray}
where
\begin{equation}
\kappa =\frac{\overline{U}_{13}^{(2)}}{\overline{U}_{11}^{(2)}} \ ,
\ \ \ \ \ 
\xi =\frac{ \overline{U}_{13}^{(3)}}{\overline{U}_{13}^{(2)}}  .
\end{equation}
Note that the proton decay due to the exchange of squarks 
$\widetilde{d_i}$
is forbidden in the limit of $\xi\rightarrow 0$, while the
radiatively-induced neutrino masses do not become zero even if
$\xi\rightarrow 0$.

\section{How to suppress the proton decay} 
\label{sec:3}

First, we discuss the parameters in the doublet sector.
We assume 
\begin{equation}
(m_-^{(2)})^2 \gg m_{SB}^2 \gg (m_+^{(2)})^2 \ .
\end{equation}
Then, we obtain the following approximate relations:
\begin{equation}
(m_1^{(2)})^2 \simeq (m_+^{(2)})^2 \left(
1 -\frac{m_{SB}^2}{(m_-^{(2)})^2}\cos^2
\alpha^{(2)}\right)
 \sim M_W^2 \ ,
\end{equation}
\begin{equation}
(m_2^{(2)})^2 \simeq (m_-^{(2)})^2 + m_{SB}^2 \sim M_I^2 \ ,
\end{equation}
\begin{equation}
\tan 2\theta_u^{(2)} = \frac{2 m_{SB} m_+^{(2)} \cos\alpha^{(2)}}{
(m_-^{(2)})^2+m_{SB}^2-(m_+^{(2)})^2} \simeq 
2\frac{m_{SB}m_+^{(2)} }{(m_-^{(2)})^2} \cos\alpha^{(2)}  \ ,
\end{equation}
\begin{equation}
s_{13}^{(2)} \equiv \overline{U}_{13}^{(2)} =  
\frac{m_{SB} \sin\alpha^{(2)}}{
\sqrt{(m_-^{(2)})^2 +m_{SB}^2 \sin^2\alpha^{(2)}}} 
\simeq  \frac{m_{SB}}{m_-^{(2)}} \sin\alpha^{(2)} \ ,
\end{equation}
\begin{equation}
s_{23}^{(2)}= -\sin \alpha^{(2)} \ , \ \ \ 
c_{23}^{(2)}=\cos\alpha^{(2)} \ ,
\end{equation}
\begin{equation}
s_{12}^{(2)} \simeq 1-\frac{1}{2}\left(\frac{m_{SB}}{m_-^{(2)}}\right)^2
 \cos^2\alpha^{(2)}
 \ , \ \ \ 
c_{12}^{(2)} \simeq  \frac{m_{SB}}{m_-^{(2)}} \cos\alpha^{(2)}
\ .
\end{equation}

Since the up-quark mass matrix $M_u$ is given by
$(M_u)_{ij} = c_u^{(2)} (Y_u)_{ij} v_u$,
where $c_u^{(2)}=\cos\theta_u^{(2)}$  
and $v_u=v \sin\beta $ ($v=174$ GeV), 
the constraint \cite{tanb} $\tan\beta >1.5$ ($\sin\beta>0.83$) 
in the conventional model from the perturbative calculability
corresponds to the constraint
\begin{equation}
c_u^{(2)} \sin\beta >0.83 \ ,
\end{equation}
which is reasonably satisfied because
the value of $c_u^{(2)}$ is given by
\begin{equation}
c_u^{(2)} \simeq \sqrt{1-\left(\frac{m_{SB} m_+^{(2)}
\cos\alpha^{(2)}}{(m_-^{(2)})^2}\right)^2 } \simeq 1 \ .
\end{equation}

On the other hand, since the down-quark mass matrix
$M_d$ is given by $(M_d)_{ij} = \overline{U}_{11}^{(2)} Y_{d ij} v_d$
($i=2,3$), where $v_d=v \cos\beta $ and
\begin{equation}
\overline{U}_{11}^{(2)} = c_{12}^{(2)} c_{13}^{(2)} \simeq
\frac{m_{SB}}{m_-^{(2)}} \cos\alpha^{(2)} \ ,
\end{equation}
the constraint \cite{tanb} $\tan\beta <60$ (i.e. $\cos\beta >0.017$) 
in the conventional model puts a constraint
$({m_{SB}}/{m_-^{(2)}}) \cos\alpha^{(2)} > {0.017}/{\cos\beta}$,
which leads to 
\begin{equation}
\frac{m_{SB}}{m_-^{(2)}} \cos\alpha^{(2)} > 0.031 \ ,
\end{equation}
where we have used the lower limit of
$\tan\beta$, $\tan\beta \simeq 1.5$.
A mass value of $\overline{H}_2$ smaller than 
$m_2^{(2)}\sim 10^{13}$ GeV cannot be accepted 
because such a small value spoils the 
coincidence of the gauge coupling constants at $\mu=M_X$.
{}From the relation (3.3), we must consider 
\begin{equation}
m_-^{(2)} \simeq m_2^{(2)} \agt 10^{14}\ {\rm GeV}\ .
\end{equation}
Therefore, a too-small value of $m_{SB}$ is not
acceptable in the present model:
\begin{equation}
m_{SB} \geq \frac{1}{\cos\alpha^{(2)}}\times 3\times 10^{12}
\ {\rm GeV}\ .
\end{equation}

The parameter values in the triplet sector are
sensitive to the proton decay.
For example, the proton decay due to the exchange of
squark $\widetilde{d}$ is proportional to 
\begin{equation}
\lambda_{1ij}^{(2,3)}\lambda_{1kl}^{(3,3)} \simeq
\xi \kappa^2 \left( \frac{m_b}{v\cos\beta}\right)^2 |V_{ub}|^2 \ ,
\end{equation}
which must be smaller than $(M_{SUSY}/M_X)^2 \sim 10^{-26}$.
Since $(m_b/v \cos\beta)^2\sim 10^{-3}$, $|V_{ub}|^2 \sim 10^{-5}$
and
\begin{equation}
\xi = \frac{ s_{13}^{(3)} }{ s_{13}^{(2)} }
\simeq \frac{ m_-^{(2)} \sin\alpha^{(3)} }{
m_-^{(3)} \sin\alpha^{(2)} }
\sim \frac{M_I}{M_X} \frac{m_5/M_X}{m_5/M_W} \sim 10^{-16}
 \ ,
\end{equation}
where we have used the relations (2.3) and the values
$m_-^{(2)}\simeq M_I \sim 10^{14}$ GeV and 
$m_-^{(3)}\simeq M_X \sim 10^{16}$ GeV, we can estimate 
the value of (3.14) as
\begin{equation}
\lambda_{1ij}^{(2,3)}\lambda_{1kl}^{(3,3)} 
\sim \kappa^2 \times 10^{-24}  \ .
\end{equation}
Since the parameter $\kappa$ is given by
\begin{equation}
\kappa = \frac{U_{13}^{(2)}}{U_{11}^{(2)} }
=\frac{\tan\theta_{13}^{(2)}}{c_{12}^{(2)}}
\simeq \tan\alpha^{(2)} \simeq \frac{m_5}{m_+^{(2)}} \ ,
\end{equation}
if we suppose $m_5 \sim 10$ GeV (i.e. $\kappa \sim 10^{-1}$), 
the proton decay due to the exchange of $\widetilde{d_3}$ 
can barely be suppressed.

On the other hand,
the proton decay due to the dimension-5 operator
is proportional to a factor
\begin{equation}
K \equiv  \frac{1}{m_1^{(3)} } c_u^{(3)}
\overline{U}_{11}^{(3)} 
+ \frac{1}{m_{2}^{(3)} } s_u^{(3)}
\overline{U}_{12}^{(3)}  \ .
\end{equation}
The value of $K$ takes a minimum at $m_-^{(3)}=m_+^{(3)}$.
Therefore, we investigate the case
\begin{equation}
(m_+^{(3)})^2 =(m_-^{(3)})^2 \gg m_{SB}^2  \ ,
\end{equation}
which have already been assumed in the derivation of (3.15).
Then, we can get the following approximate relations:
\begin{equation}
(m_1^{(3)})^2 \simeq (m_+^{(3)})^2 - m_{SB} m_+^{(3)} \ ,
\ \ \ 
(m_2^{(3)})^2 \simeq (m_+^{(3)})^2 + m_{SB} m_+^{(3)} \ , 
\end{equation}
\begin{equation}
\tan 2\theta_u^{(3)} \simeq 2 \frac{m_+^{(3)}}{m_{SB}} \ , \  \ 
{\rm i.e.} \ \ 
\cos 2\theta_{u}^{(3)} \simeq \frac{m_{SB}}{2 m_+^{(3)}} \ ,
\end{equation}
\begin{equation}
\overline{U}^{(3)} \simeq \left(
\begin{array}{ccc}
c_{12}^{(3)} & s_{12}^{(3)} & s_{13}^{(3)} \\
-s_{12}^{(3)} & c_{12}^{(3)} & s_{23}^{(3)} \\
s_{23}^{(3)}s_{12}^{(3)} & 
-s_{23}^{(3)}c_{12}^{(3)} & 1 \\
\end{array} \right) \ ,
\end{equation}
where
\begin{equation}
\cos 2\theta_{12}^{(3)} \simeq \frac{m_{SB}}{2 m_\pm^{(3)}} \ ,
\end{equation}
\begin{equation}
s_{13}^{(3)} \simeq \frac{m_{SB}}{m_-^{(3)}}\sin\alpha^{(3)}
\simeq \frac{m_{SB}}{m_-^{(3)}} \frac{m_5}{m_+^{(3)}}
\sim 10^{-19} \ ,
\end{equation}
\begin{equation}
s_{23}^{(3)} =- \sin\alpha^{(3)} \simeq - \frac{m_5}{m_+^{(3)}}
\sim -10^{-15} \ .
\end{equation}
Therefore, the factor $K$ is estimated as
\begin{equation}
K \simeq -\frac{m_{SB}}{(m_+^{(3)})^2} \sim - 10^{-4} \frac{1}{M_X} \ .
\end{equation}
In order to suppress the proton decay due to
the dimension-5 operator, it is better to take the value 
of $m_{SB}$ as low as possible.

For example, the numerical values without approximation are
as follows: for the input values
\begin{eqnarray}
m_{SB} &=& 4\times 10^{12} \ {\rm GeV} \ , 
\ \ \ \ m_5 = 2\times 10^1\ {\rm GeV} \ , \nonumber \\
m_+^{(2)}& =& 2\times 10^2 \  {\rm GeV} \ , \ \ \ \ 
 m_-^{(2)} = 1 \times 10^{14} \  {\rm GeV} \ ,  \nonumber \\
 m_+^{(3)} &=& 5\times 10^{16} \  {\rm GeV} \ , \ \ \ \ 
m_-^{(3)} = 5 \times 10^{16} \  {\rm GeV} \ ,  \\
 \sin\alpha^{(2)} &=& 0.1 \ ,  \ \ \ \ 
\sin\alpha^{(3)}=4\times 10^{-16} \ , \nonumber
\end{eqnarray}
we obtain 
\begin{equation}
m_1^{(2)} = -2\times 10^2\  {\rm GeV} \ , \ \  
m_2^{(2)} = 1.0 \times 10^{14} \,  {\rm GeV} \ ,
\end{equation}
\begin{equation}
U^{(2)} = \left(
\begin{array}{cc}
1 & 7.9\times 10^{-14} \\
-7.9\times 10^{-14} & 1
\end{array} \right) \ ,
\end{equation}
\begin{equation}
\overline{U}^{(2)} = \left(
\begin{array}{ccc}
0.040 & 0.999 & 0.004\\
-0.994 & 0.040 & -0.100 \\
-0.100 & 0.9\times 10^{-16} & 0.995
\end{array} \right) \ ,
\end{equation}
\begin{equation}
m_1^{(3)} = -5.0 \times 10^{16}\, {\rm GeV} \ , \ \  
m_2^{(3)} = 5.0 \times 10^{16} \,  {\rm GeV} \ ,
\end{equation}
\begin{equation}
U^{(3)} = \left(
\begin{array}{cc}
0.707 & 0.707 \\
-0.707 & 0.707
\end{array} \right) \ ,
\end{equation}
\begin{equation}
\overline{U}^{(3)} = \left(
\begin{array}{ccc}
0.707 & 0.707 & 3.2\times 10^{-20}\\
-0.707 & 0.707 & -4.0\times 10^{-16} \\
-2.8\times 10^{-16} & 2.8 \times 10^{-16} & 1.000
\end{array} \right) \ ,
\end{equation}
\begin{equation}
\overline{U}_{11}^{(2)}=0.040 \ ,\ \ \ K=-3.2 \times 10^{-5}/M_X \ ,
\ \ \ \kappa = 0.10 \ , \ \ \xi =0.8 \times 10^{-17} \ .
\end{equation}
Therefore, these parameter values are harmless for the proton decay.

\begin{figure}[hb]
\unitlength=1cm
\begin{picture}(16.6,4)
\thicklines
%
%
\put(0.5,1){\line(1,0){2}}
\put(1.5,1){\vector(1,0){0}}
\put(0.8,1.3){$\nu_j$}
\put(2.5,1){\circle*{0.2}}
\multiput(2.5,1)(0.5,0){8}{\line(1,0){0.3}}
\put(3.5,1){\vector(1,0){0}}
\put(3,0.3){$\tilde{e}_R$}
\put(4.5,1){\circle*{0.2}}
\put(5.5,1){\vector(1,0){0}}
\put(5.5,0.3){$\tilde{e}_L$}
\put(6.5,1){\circle*{0.2}}
\put(4.2,0.3){$\widetilde{M}^{2}_{eLR}$}
\put(6.5,1){\line(1,0){2}}
\put(8,1){\vector(-1,0){0}}
\put(8,1.3){$\nu_i^c$}
\put(4.5,1){
\qbezier(-2,0)(-2.01,0.35)(-1.88,0.68)
\qbezier(-1.88,0.68)(-1.78,1.03)(-1.53,1.29)
\qbezier(-1.53,1.29)(-1.3,1.55)(-1,1.73)
\qbezier(-1,1.73)(-0.69,1.9)(-0.35,1.97)
\qbezier(-0.35,1.97)(0,2.03)(0.35,1.97)
\qbezier(2,0)(2.01,0.35)(1.88,0.68)
\qbezier(1.88,0.68)(1.78,1.03)(1.53,1.29)
\qbezier(1.53,1.29)(1.3,1.55)(1,1.73)
\qbezier(1,1.73)(0.69,1.9)(0.35,1.97)
}
\put(4.4,3){\circle*{0.2}}
\put(4.2,3.3){$M_e$}
\put(2.3,2.3){$e_L$}
\put(6.6,2.3){$e_R$}
%
%
\put(8.8,1){\line(1,0){2}}
\put(9.8,1){\vector(1,0){0}}
\put(9.1,1.3){$\nu_j$}
\put(10.8,1){\circle*{0.2}}
\multiput(10.8,1)(0.5,0){8}{\line(1,0){0.3}}
\put(11.8,1){\vector(1,0){0}}
\put(11.3,0.3){$\tilde{d}_R$}
\put(12.8,1){\circle*{0.2}}
\put(13.8,1){\vector(1,0){0}}
\put(13.8,0.3){$\tilde{d}_L$}
\put(14.8,1){\circle*{0.2}}
\put(12.5,0.3){$\widetilde{M}^{2}_{dLR}$}
\put(14.8,1){\line(1,0){2}}
\put(16.3,1){\vector(-1,0){0}}
\put(16.3,1.3){$\nu_i^c$}
\put(12.8,1){
\qbezier(-2,0)(-2.01,0.35)(-1.88,0.68)
\qbezier(-1.88,0.68)(-1.78,1.03)(-1.53,1.29)
\qbezier(-1.53,1.29)(-1.3,1.55)(-1,1.73)
\qbezier(-1,1.73)(-0.69,1.9)(-0.35,1.97)
\qbezier(-0.35,1.97)(0,2.03)(0.35,1.97)
\qbezier(2,0)(2.01,0.35)(1.88,0.68)
\qbezier(1.88,0.68)(1.78,1.03)(1.53,1.29)
\qbezier(1.53,1.29)(1.3,1.55)(1,1.73)
\qbezier(1,1.73)(0.69,1.9)(0.35,1.97)
}
\put(12.7,3){\circle*{0.2}}
\put(12.5,3.3){$M_d$}
\put(10.6,2.3){$d_L$}
\put(14.9,2.3){$d_R$}
\end{picture}
\caption{Radiative generation of neutrino Majorana mass}
\label{fig:numass}
\end{figure}
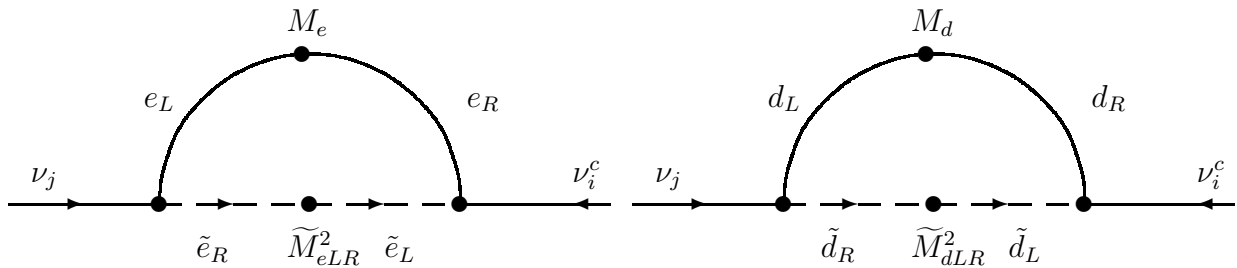

\section{Radiatively induced neutrino mass matrix} 
\label{sec:4}

In a SUSY GUT scenario, there are many origins of the 
neutrino mass generations \cite{Hall}.
For example, the sneutrinos $\widetilde{\nu}_{iL}$ can
have VEVs $\langle \widetilde{\nu}\rangle \neq  0$, and  
the neutrinos $\nu_{Li}$ acquire
their masses thereby \cite{Diaz}.
In the present model, there is an $R$-parity 
violating bilinear term $\overline{5}_{L(+)} H_{(+)}$,
while there is no $\overline{H}_{(-)} H_{(+)}$ term
(the so-called $\mu$-term).
In the physical field basis (the basis on which the
mass matrix (2.5) is diagonal),  the so-called $\mu$-term, 
$m_1 \overline{H}_1 H_1$, appears,
while the $\overline{5}_L H_1$ term is absent.
Therefore, in the present model, the sneutrinos cannot
have VEVs $\langle \tilde{\nu}_i \rangle$ at tree level.
The VEVs $\langle \tilde{\nu}_i \rangle \neq 0$ will appear
only through the renormalization group equation (RGE) effect.
The contribution highly depends on an explicit model
of the SUSY symmetry breaking.
Since the purpose of the present paper is to investigate
a general structure of the radiative neutrino masses,
for the moment, we confine ourselves to discussing 
possible forms of the radiative neutrino mass matrix .

The radiative neutrino mass matrix 
$M_\nu^{rad}$ is given by 
\begin{equation}
M_\nu^{rad} = M_\nu^e + M_\nu^d \ ,
\end{equation}
where $M_\nu^e$ is generated
by the interactions $\nu_L e_L \widetilde{e}_R^c$ 
and $\nu_L \widetilde{e}_L e_R^c$ (i.e. by the charged lepton loop) 
and $M_\nu^d$ is generated by $\nu_L d_R^c \widetilde{d}_L$ and 
$\nu_L \widetilde{d}_R^cd_L$ (i.e. by the down-quark loop). 
We assume that the contributions from Zee-type diagrams due to 
$\overline{H}^+ \leftrightarrow\widetilde{e}^+_{R}$ mixing 
is negligibly small because the term 
$\overline{H}\,\overline{H}\, 10_L$ must be
not $\overline{H}_1\overline{H}_1 10_L$, but 
 $\overline{H}_1\overline{H}_2 10_L$ 
(recall that only the field $\overline{H}_1$ has the VEV in the
present model).

We consider the radiative diagram with 
$(\nu_L)_j \rightarrow (e_R)_l + (\widetilde{e}_L^c)_n$ and
$(e_L)_k + (\widetilde{e}_L^c)_m \rightarrow (\nu_L^c)_i$.
The contributions $(M_\nu^e)_{ij}$ from the charged lepton loop are,
except for the common factors, given as follows:
\begin{equation}
(M_\nu^e)_{ij}= (\lambda_{1km} \delta_{i1} -\lambda_{1im}\delta_{k1})
(\lambda_{1jl} \delta_{n1} -\lambda_{1nl}\delta_{j1}) (M_e)_{kl}
(\widetilde{M}_{eLR}^{2})_{nm} + (i \leftrightarrow j) \ ,
\end{equation}
where $M_e$ and $\widetilde{M}_{eLR}^2$ are charged-lepton and 
charged-slepton-$LR$ mass matrices, respectively.
Here and hereafter, we will drop the common factor in
$(M_\nu^{rad})_{ij}$, because we have an interest only in the
relative structure of the matrix elements $(M_\nu^{rad})_{ij}$.
Since, as usual, we assume that the structure of 
$\widetilde{M}_{eLR}^2$ 
is proportional to that of $M_e$, we obtain
\begin{eqnarray}
(M_\nu^e)_{ij}&=& \lambda_{1im} \lambda_{1jl} (M_e)_{1l} (M_e)_{1m}
+\delta_{i1} \delta_{j1} \lambda_{1km} \lambda_{1nl}
(M_e)_{kl} (M_e)_{nm} \ \nonumber \\
& &
-\delta_{i1} \lambda_{1jl} \lambda_{1km} (M_e)_{1m} (M_e)_{kl}
-\delta_{j1} \lambda_{1il} \lambda_{1km} (M_e)_{1m} (M_e)_{kl} \ .
\end{eqnarray}
Since $\lambda^e_{1ij} \equiv \lambda_{1ij}^{(2,2)}=
\kappa (1-\delta_{i1})(M_e^\dagger)_{ji}$ 
from the expression (2.32),
we obtain the contribution from the charged lepton loop:  
\begin{equation}
M_\nu^e =H_e^T S_1 H_e - S_1 H_eH_e -H_e^T H_e^T S_1 
+ S_1 {\rm Tr}(H_eH_e) \ .
\end{equation}
where we have dropped the common factor 
$\kappa$,
and the Hermitian matrix $H_e$ and the rank-1 matrix $S_1$ are 
defined by
\begin{equation}
H_e=M_e M_e^\dagger \ ,
\end{equation}
\begin{equation}
S_1 = \left(
\begin{array}{ccc}
1 & 0 & 0 \\
0 & 0 & 0 \\
0 & 0 & 0
\end{array} \right) \ .
\end{equation}

Similarly, we can obtain the contributions from the down-quark loop.
{}From the expression (2.33), we denote 
$\lambda_{1ij}^d \equiv \lambda_{1ij}^{(2,3)}$ as
\begin{equation}
\lambda_{1ij}^d =\kappa [\rho\delta_{1i} +(1-\delta_{i1})]
(M_d^\dagger)_{ij}
\ ,
\end{equation}
where 
\begin{equation}
\rho = \frac{1}{1-\xi\kappa U_{31}^{(2)}/U_{33}^{(3)}}
\simeq 1 \ .
\end{equation}
Then, we obtain
\begin{equation}
M_\nu^d =H_d S_1 H_d^T - S_1 H_d^T H_d^T - H_dH_d S_1 
+ S_1 {\rm Tr}(H_dH_d) \ ,
\end{equation}
where
\begin{equation}
H_d=M_d^\dagger M_d \ .
\end{equation}
Note that the result (4.9) is independent of the value of $\rho$.

The field $\overline{5}_{L(+)1}$ defined in Eq.~(2.2) 
does not mean the
observed first-generation field $(d^c,\nu,e)_L$ (and its SUSY
partner).
The forms of $M_\nu^e$ and $M_\nu^d$ on the general basis are 
given by
\begin{equation}
M_\nu^e =H_e^T S H_e - S H_eH_e - H_e^T H_e^T S + S{\rm Tr}(H_eH_e) \ ,
\end{equation}
\begin{equation}
M_\nu^d =H_d S H_d^T - S H_d^T H_d^T - H_dH_d S + S{\rm Tr}(H_dH_d) \ ,
\end{equation}
where $S$ is an arbitrary rank-1 matrix $S=U_5^T S_1 U_5$,
which is given by the rebasing $\overline{5}_i 
\rightarrow \overline{5}'_i =(U_5^\dagger \overline{5} )_i$.

It is convenient to investigate the form $M_\nu^{rad}$ on the basis
on which the charged lepton mass matrix $M_e$ is diagonal:
$H_e=D_e^2={\rm diag}(m_{e1}^2, m_{e2}^2, m_{e3}^2) 
\equiv {\rm diag}(m_e^2, m_\mu^2, m_\tau^2)$.
Then, the matrix $M_\nu^e$ is given by
\begin{equation}
M_\nu^e = \left(
\begin{array}{ccc}
S_{11} (  m_{e2}^4 + m_{e3}^4 ) & 
S_{12} ( m_{e3}^4 +m_{e1}^2 m_{e2}^2) &
S_{13} ( m_{e2}^4 +m_{e1}^2 m_{e3}^2) \\
S_{21} ( m_{e3}^4 +m_{e1}^2 m_{e2}^2) &
S_{22} (  m_{e3}^4 + m_{e1}^4 ) &
S_{23} ( m_{e1}^4 +m_{e2}^2 m_{e3}^2) \\
S_{31} ( m_{e2}^4 +m_{e1}^2 m_{e3}^2) &
S_{32} ( m_{e1}^4 +m_{e2}^2 m_{e3}^2) &
S_{33} (  m_{e1}^4 + m_{e2}^4 )
\end{array} \right) \ .
\end{equation}
On the basis with $H_e=D_e^2$, since the matrix $H_d$ can 
expressed as $H_d = U D_d^2 U^T$, where $U \equiv U_R^d$, 
the contribution $M_\nu^d$ is expressed as
\begin{eqnarray}
& M_\nu^d =U ({M}_\nu^d)' U^T \ , \\
& ({M}_\nu^d)' = D_d^2 {S}' D_d^2
-{S}' D_d^4 - D_d^4 {S}'
-{S}' {\rm Tr}D_d^4 \ , \\
& {S}' = U^\dagger S U^* \ .
\end{eqnarray}
Here, $({M}_\nu^d)'$ is again given by the
expression similar to (4.13):
\begin{equation}
({M}_\nu^d)' = \left(
\begin{array}{ccc}
{S}'_{11} (  m_{d2}^4 + m_{d3}^4 ) & 
{S}'_{12} ( m_{d3}^4 +m_{d1}^2 m_{d2}^2) &
{S}'_{13} ( m_{d2}^4 +m_{d1}^2 m_{d3}^2) \\
{S}'_{21} ( m_{d3}^4 +m_{d1}^2 m_{d2}^2) &
{S}'_{22} (  m_{d3}^4 + m_{d1}^4 ) &
{S}'_{23} ( m_{d1}^4 +m_{d2}^2 m_{d3}^2) \\
{S}'_{31} ( m_{d2}^4 +m_{d1}^2 m_{d3}^2) &
{S}'_{32} ( m_{d1}^4 +m_{d2}^2 m_{d3}^2) &
{S}'_{33} (  m_{d1}^4 + m_{d2}^4 )
\end{array} \right) \ .
\end{equation}
The mass ratios $m_s^2/m_b^2 \simeq 7.02\times
10^{-4}$ and $m_\mu^2/m_\tau^2 \simeq 3.43\times
10^{-3}$ at $\mu=M_X$ are negligibly small
compared with $\Delta m^2_{solar}/\Delta m^2_{atm}
\sim 10^{-2}$, so that when we neglect the terms
with $m_{e1}^2/m_{e3}^2$, $m_{e2}^2/m_{e3}^2$, 
$m_{d1}^2/m_{d3}^2$ and $m_{d2}^2/m_{d3}^2$,  
we can approximate (4.13) and (4.17) as
\begin{equation}
M_\nu^e \simeq m_\tau^4 \left(
\begin{array}{ccc}
S_{11} & S_{12} & 0 \\
S_{21} & S_{22} & 0 \\
0 & 0 & 0
\end{array} \right) =  m_\tau^4 PSP \ , \ \ 
({M}_\nu^d)' \simeq  m_b^4 \left(
\begin{array}{ccc}
{S}'_{11}  & {S}'_{12}  & 0 \\
{S}'_{21}  & {S}'_{22}  & 0 \\
0 & 0 & 0
\end{array} \right) = m_b^4 PS'P \ ,
\end{equation}
where $P$ is defined as 
\begin{equation}
P =  {\rm diag}(1, 1, 0) \ .
\end{equation}
Therefore, we can express the neutrino mass matrix 
$M_\nu^{rad}$ 
as the following form:
\begin{equation}
M_\nu^{rad} = m_0 \left( PSP + k U \cdot P{U^\dagger S U^*}P
\cdot U^T \right) ,
\end{equation}
where $k$ is given by $k \simeq (m_b/m_\tau)^2$ and 
$m_0$ will be given later [in Eq.~(4.22)].
The matrix $S$ is a rank-1 matrix, so that $PSP$,
$U^\dagger S U^*$, and $U (P{U^\dagger S U^*}P)U^T$ 
are also rank-1 matrices.
In other words, the radiative neutrino mass matrix $M_\nu^{rad}$ has
the form which is described by two rank-1 matrices:
\begin{equation}
M_\nu^{rad} = m_0 \left(
\begin{array}{ccc}
g_1^2 & g_1 g_2 & 0 \\
g_2 g_1 & g_2^2 & 0 \\
0 & 0 & 0
\end{array} \right) + m_0 k \left(
\begin{array}{ccc}
f_1^2 & f_1 f_2 & f_1 f_3 \\
f_2 f_1 & f_2^2 & f_2 f_3 \\
f_3 f_1 & f_3 f_2 & f_3^2
\end{array} \right) \ .
\end{equation}

So far, we have not discussed the absolute magnitude of
the neutrino mass matrix $M_\nu^{rad}$. When we assume 
$m^2 (\widetilde{e}_R) \equiv m^2 (\widetilde{e}_{R3})
\simeq m^2 (\widetilde{e}_{R2}) \simeq  m^2 
(\widetilde{e}_{R1})$ and $m^2 (\widetilde{e}_L) \equiv
m^2 (\widetilde{e}_{L3}) \simeq m^2 (\widetilde{e}_{L2})
\simeq m^2 (\widetilde{e}_{L1})$ and the rank-1 matrix
$S$ is normalized as $ {\rm Tr} (SS^\dagger)=1$, the 
coefficient $m_0$ in the expression (4.20) is given by
\begin{equation}
m_0 = \frac{1}{16 \pi^2} \kappa^{2} 
\frac{m^{(2)}_{1} m^4_\tau}{v^2} 
F(m^{2} (\widetilde{e}_R),m^{2} (\widetilde{e}_L) )
\ ,
\end{equation}
where
\begin{equation}
F(m^{2}_R,m^{2}_L)
=\frac{1}{m^{2}_{R}-m^{2}_{L}} \ln 
\frac{m^{2}_{R}}{m^{2}_{L}}
 \ .
\end{equation}
If $F(m^{2} (\widetilde{e}_R),m^{2} (\widetilde{e}_L) )
\simeq F(m^{2} (\widetilde{d}_R),m^{2} (\widetilde{d}_L))$,
the factor $k$ is given by $k \simeq (m_{b}/m_{\tau})^4=
8.6$. However, in the present paper, we regard $k$ as a 
free parameter. 
By using $1/16 \pi^{2} = 6.33 \times 10^{-3}$, 
$m^{(2)}_{1}\equiv m(H^{(2)}_1)=2 \times 10^{2} \ {\rm GeV}$, 
$m_\tau (m_Z) =1.75$ GeV, $v=174$ GeV and $\tan\beta =1.5$, 
we obtain
\begin{equation}
m_0 \simeq 1.9 \kappa^{2} F \ {\rm eV}
\ ,
\end{equation}
where $F$ is the value of $F(m^{2} (\widetilde{e}_R) ,
m^{2} (\widetilde{e}_{L}))$ in the unit of {\rm TeV}. 
If the neutrino mass matrix $M_\nu$ is dominated by
the radiative mass terms $M_\nu^{rad}$ and 
we wish that the largest one of $m_{\nu i}$ is of the order
of $\sqrt{\Delta m^2_{atm}} \simeq 0.05 \ {\rm eV}$, the 
value $\kappa \sim 10^{-1}$ is favorable.


\section{Possible form of $M_\nu$} 
\label{sec:5}

The neutrino mass matrix in the present model is
given by
\begin{equation}
M_{\nu} = M_\nu^{rad} + M_{\tilde{\nu}} \ .
\end{equation}
The contribution  $M_{\tilde{\nu}} $ from 
$\langle \tilde{\nu}_i \rangle \neq 0$
is estimated as follows.
Since the mass matrix for $(\nu_1, \nu_2 , \nu_3 , 
\widetilde{W}^0)$ (except
for the radiative masses) is given by
\begin{equation}
\left(
\begin{array}{cccc}
0 & 0 & 0 & \frac{1}{2}g v_1 \\
0 & 0 & 0 & \frac{1}{2}g v_2 \\
0 & 0 & 0 & \frac{1}{2}g v_3 \\
\frac{1}{2}g v_1 & \frac{1}{2}g v_2 & \frac{1}{2}g v_3 
& M_{\widetilde{W}}
\end{array} \right) \ ,
\end{equation}
where, for simplicity, we have dropped the elements for
$\widetilde{B}^0$, we obtain
\begin{equation}
M_{\tilde{\nu}} 
\simeq 
- \frac{1}{4}g^2
\left(
\begin{array}{c}
v_1  \\
v_2  \\
v_3 
\end{array} \right) 
(M_{\widetilde{W}})^{-1}
(v_1 \ v_2 \ v_3)
= -\frac{g^2}{4 M_{\widetilde{W}}}
\left(
\begin{array}{ccc}
v^2_1 & v_1 v_2 & v_1 v_3 \\
v_1 v_2 & v^2_2 & v_2 v_3 \\
v_1 v_3 & v_2 v_3 & v_3^2 
\end{array} \right) 
\ ,
\end{equation}
under the seesaw approximation.
Note that the matrix $M_{\tilde{\nu}}$ is a rank-1 matrix.

As noted in Sec.~IV, the sneutrinos in the present model
cannot have non-zero VEVs at the tree level.
The VEVs $\langle \tilde{\nu}_i \rangle \neq 0$ can appear
only through RGE effect.
Therefore, the magnitudes of $v_i\equiv \langle \tilde{\nu}_i
\rangle$ are highly dependent on a model of the SUSY breaking
and the RGE effect.
In the previous section, we have estimate the magnitudes of
the radiative masses, i.e. in Eq.~(4.24).
In the present paper, we interest only in the form of the
neutrino mass matrix $M_\nu$.
Therefore, we do not discuss the explicit symmetry
breaking mechanism and the absolute magnitudes of
$\langle \tilde{\nu}_i \rangle $.

Since the form of the flavor symmetry breaking in the
present model is described by the rank-1 matrix $S$,
it is likely that the structure of the rank-1 matrix
$M_{\tilde{\nu}}$ is also given by the matrix $S$, i.e.
$M_{\tilde{\nu}} = m_0 r S$,
where the factor $r$ denotes a relative ratio of 
$M_{\tilde{\nu}}$ to $M_\nu^{rad}$.
Then, the neutrino mass matrix $M_\nu$ is expressed as
\begin{equation}
M_\nu = m_0 \left(r S + PSP + 
k U \cdot P{U^\dagger S U^*}P \cdot U^T  \right) ,
\end{equation}
and 
\begin{equation}
M_\nu = m_0 \left(
\begin{array}{ccc}
g_1^2 (1+r) & g_1 g_2 (1+r) & g_1 g_3 r \\
g_2 g_1 (1+r) & g_2^2 (1+r) & g_2 g_3 r \\
g_3 g_1 r & g_2 g_3 r & g_3^2 r
\end{array} \right) + m_0 k \left(
\begin{array}{ccc}
f_1^2 & f_1 f_2 & f_1 f_3 \\
f_2 f_1 & f_2^2 & f_2 f_3 \\
f_3 f_1 & f_3 f_2 & f_3^2
\end{array} \right) \ .
\end{equation}
correspondingly to the expressions (4.20) and (4.21), 
respectively.

It is interesting to consider a case that the neutrino mass
matrix $M_\nu$ is dominated by the radiative masses $M_\nu^{rad}$.
Or, we also interests in a case with $S$ which satisfies
the relation $PSP=S$ [a case with $g_3=0$ in the expression (5.5)].
Then, since the neutrino mass matrix $M_\nu$ is still given
by the form (4.20) [(4.21)], i.e. by the sum of two rank-1 matrices, 
it gives det$M_\nu =0$, 
so that one of the eigenvalues of $M_\nu$ is zero.
Therefore, we can consider the following two cases:
\begin{equation}
m_{\nu 1} = 0, \ \ m_{\nu 2}= m_0 \varepsilon , \ \ m_{\nu 3}= m_0  ,
\end{equation}
for a normal hierarchy model, and
\begin{equation}
m_{\nu 1} = \frac{1}{2} m_0 (1-\varepsilon^2), \ \ 
m_{\nu 2} = \frac{1}{2} m_0 (1+\varepsilon^2), \ \ 
m_{\nu 3}= 0 ,
\end{equation}
for an inverse hierarchy model, where
\begin{equation}
\varepsilon \simeq \sqrt{|R|} \ ,
\end{equation}
\begin{equation}
R \equiv \frac{\Delta m_{solar}^2}{\Delta m_{atm}^2} \ .
\end{equation}
The inverse hierarchy case with the eigenvalues 
$m_{\nu 1} = -(1/2) m_0 (1-\varepsilon^2)$, 
$m_{\nu 2} = (1/2) m_0 (1+\varepsilon^2)$ and
$m_{\nu 3}= 0 $ is ruled out in the present model,
because the case gives $\sum_i m_{\nu i} = m_0 \varepsilon^2$,
while the mass matrix (4.21) gives Tr$M_\nu= m_0
(\sum_i g_i^2 + k \sum_i f_i^2)$, so that $k$ must be
negative to give a small value of Tr$M_\nu$.
However, it is unlikely that the contributions 
$M_\nu^e$ and $M_\nu^d$ have opposite signs each other.

In Appendix B, we will show that the form (4.20) can always
contain parameter values which lead to a nearly bimaximal 
mixing
\begin{equation}
U_\nu =  \left(
\begin{array}{ccc}
c_{12}\sqrt{1-\varepsilon_{13}^2} & s_{12}\sqrt{1-\varepsilon_{13}^2} 
& \varepsilon_{13} \\
-\frac{s_{12}-c_{12}\varepsilon_{13}}{\sqrt2} &
\frac{c_{12}+s_{12}\varepsilon_{13}}{\sqrt2} &
 -\frac{\sqrt{1-\varepsilon_{13}^2}}{\sqrt2} \\
-\frac{s_{12}+c_{12}\varepsilon_{13}}{\sqrt2} &
\frac{c_{12}-s_{12}\varepsilon_{13}}{\sqrt2} &
\frac{\sqrt{1-\varepsilon_{13}^2}}{\sqrt2} \\
\end{array} \right) \ ,
\end{equation}
where $c_{12} \sim s_{13}$ and 
$\varepsilon^2_{12} \ll 1$, and to the ratio 
$R \equiv \Delta m^2_{21} / \Delta m^2_{32} \sim 10^{-2}$.
However, even if we assume the dominance of $M_\nu^{rad}$
in $M_\nu$, 
since the expression (4.20) has many free parameters,
we cannot give any predictions for the neutrino phenomenology
unless we put a further ansatz for the flavor symmetry. 

In Appendix C, we demonstrate a simple example of $S$ which
satisfies the relation $PSP=S$. 
The model can lead to a successful description of the observed
neutrino masses and mixings \cite{solar,kamland,atm}.
However, this is merely one of the examples. 
The systematical search for the explicit form $M_\nu$ and 
a possible flavor symmetry is be a future task.

\section{Summary} 
\label{sec:6}

In conclusion, within the framework of an SU(5) SUSY GUT
model, we have proposed a mechanism which effectively
induces $R$-parity-violating terms at $\mu < m_{SB}$. In our model,
those terms with lepton number violation are large enough to generate
neutrino Majorana masses while those with baryon number violation
are strongly suppressed so that the experimental bound of proton decay
is evaded. This is related with doublet-triplet splitting. We have
matter fields $\overline{5}_{L(+)} + 10_{L(-)}$
and two types of Higgs fields $H_{(\pm)}$ and 
$\overline{H}_{(\pm)}$, where $(\pm)$ denote the 
transformation properties under a discrete symmetry Z$_2$.
The Higgs fields $H_{(+)}$ and $\overline{H}_{(-)}$ couple
to $10_{L(-)} 10_{L(-)}$ 
and $\overline{5}_{L(+)} 10_{L(-)}$,
respectively, to make the Yukawa interactions. The Z$_2$ symmetry
is only broken by the $\mu$-term, $m_{SB} \overline{H}_{(+)} H_{(-)}$,
so that the $\overline{H}_{(-)} \leftrightarrow \overline{5}_{(+)}$
mixing is effectively induced at $\mu < m_{SB}$.
Because of the heaviness of the color triplet components of the
Higgs fields, the mixing is sizable in the $SU(2)_L$ doublet 
sector, while it is negligibly small in the $SU(3)_c$ triplet sector.

Whether the model is harmless or not for proton decay is highly
sensitive to the choice of the parameter values, especially, $m_{SB}$
and $m_5$. 
A smaller value of $m_{SB}$ gives a lighter mass for the massive Higgs 
fields $H_2$ (another one, $H_1$, corresponds to the Higgs field 
in the conventional model),  so that the case spoils the unification 
of the gauge coupling constants at $\mu=M_X$. 
On the other hand, a large value of $m_{SB}$ induces the proton
decay due to the dimension-5 operator. We have taken 
$m_{SB} \sim 10^{14} \ {\rm GeV}$. Also, a large value of $m_5$ induce the 
proton decay due to the exchange of squark $\widetilde{d}$. We have
taken $m_5 \sim 10^{1} \ {\rm GeV}$. Those parameter values can give a
reasonable magnitude of the neutrino mass. However, the choice of such
a small $m_5$ gives a small mixing between $\overline{H}_{(+)}$ and
$\overline{5}_{(+)}$, so that the case gives $\overline{U}^{(2)}_{22}
\simeq \overline{U}^{(2)}_{11}$ and $\overline{U}^{(2)}_{11} 
\overline{U}^{(2)}_{33} - \overline{U}^{(2)}_{13} 
\overline{U}^{(2)}_{31} \simeq \overline{U}^{(2)}_{11} 
\overline{U}^{(3)}_{33} - \overline{U}^{(3)}_{13} 
\overline{U}^{(2)}_{31}$. Therefore, the case with $|\alpha^{(2)}| \ll 1$
cannot give a sizable deviation from $M^T_d = M_e$. 
However, this is critical for each parameter value. 
The details are dependent on the 
explicit model, i.e. on the choice of the forms $S$ and 
$U \equiv U^d_R$. A further careful study based on an explicit model
will be required.

Anyhow, if the present scenario is working, the proton decay will be
observed in the near future, because possible parameter values are in
critical ranges for the proton decay in order to explain the quark and
lepton (charged lepton and neutrino) masses and mixings.

The present model leads to a radiatively-induced neutrino mass
matrix $M_\nu^{rad}$ which is given by sum of two rank-1 matrices
as shown in Eq.~(4.20).
The ``two" is originated in the two contributions from
charged lepton loop and down-quark loop.
The reason that each contribution takes a rank-1 matrix form is
owing to that the mixing of the matter fields $\overline{5}_{L(+)i}$
($i=1,2,3$) with the Higgs field $\overline{H}_{(-)}$ is take place
only for a linear combination $\sum b_i \overline{5}_{L(+)i}$.
The contribution $M_{\tilde{\nu}}$ from 
$\langle \tilde{\nu}_i \rangle \neq 0$ is also expressed by
a rank-1 matrix.
Then, the general form of $M_\nu$ is given by the expression (5.4)
[(5.5)].

We have investigated an interesting case that the form of
$M_{\tilde{\nu}}$ is given by a rank-1 matrix $S$, especially,
the case with $PSP=P$.
Then, the neutrino mass matrix $M_\nu$ is given by the form (4.20).
Since two rank-1 matrix model generally gives det$M_\nu=0$,
one of the eigenvalues has to be zero, so that we can consider
two types of the mass hierarchy: the normal hierarchy with
$D_\nu=m_0{\rm diag}(0,\varepsilon, 1)$, and the inverse
hierarchy with $D_\nu=(1/2)m_0{\rm diag}(1-\varepsilon^2, 
1+\varepsilon^2, 0)$, where $\varepsilon^2 \simeq \Delta m^2_{solar}/
\Delta m^2_{atm}$.
The case of the inverse hierarchy with $D_\nu=(1/2)m_0
{\rm diag}(\varepsilon^2 -1, \varepsilon^2 +1, 0)$ is ruled out.
However, even if we assume that the observed neutrino masses
and mixings are dominantly described  by the radiative neutrino mass
matrix (4.20) (or $S$ satisfies $PSP=S$), 
we cannot yet give an explicit predictions unless
we assume a further ansatz for the flavor symmetry, because we
have many free parameters in the rank-1 matrix $S$ and 
the unitary matrix $U\equiv U_R^d$.
For a flavor symmetry in the neutrino mass matrix $M_\nu$,
we have known that a $2\leftrightarrow 3$ permutation
symmetry is promising \cite{2-3sym}.
For example, a successful example given in Appendix C  
satisfies the $2\leftrightarrow 3$ symmetry.
Moreover, a possibility that the $2\leftrightarrow 3$
symmetry can be applicable to the unified description
of quark and charged lepton mass matrices has been
pointed out \cite{Koide-Nishiura}.
However, since the purpose of the present paper is
to give an $R$-parity violation mechanism (and the
radiative neutrino mass matrix $M_\nu^{rad}$ within
the framework of an SU(5) SUSY GUT without any
troubles for proton decay, we have discussed the details
no more.
It will be our next task to seek for what flavor symmetry 
is reasonable.

Nevertheless, the present model will be worth noticing.
In the present model, the coupling constants $\lambda_{ijk}$
of $\nu_{Li} e_{Lj} e_{Rk}^c$ and $\nu_{Li} d_{Rj}^c d_{Lk}$
are proportional to the mass matrices $(M_e^*)_{jk}$ and
$(M_d^\dagger)_{jk}$, respectively. 
The model will give fruitful phenomenology in flavor
violating processes.

\vspace{7mm}

\centerline{\large\bf Acknowledgments}

This work was supported by the Grant-in-Aid for
Scientific Research, the Ministry of Education,
Science and Culture, Japan (Grant Numbers 14039209, 14046217,
1474068 and 15540283).

\vspace{7mm}

\centerline{\large\bf Appendix A. General form of the mass
matrix $M$} 

\vspace{4mm}

The Z$_2$ symmetry can softly be violated not only by the terms
$\overline{H}_{(+)}{H}_{(-)}$, but also by terms $\overline{H}_{(-)}
{H}_{(+)}$ and $\overline{5}_{L(+)i}{H}_{(-)}$. The mass matrix $M$
given in Eq.~(2.5) is generally represented by
$$
M =  \left(
\begin{array}{cc}
m^{-+}_{SB} & m_- \\
m_{+} c_{\alpha} &  m^{+-}_{SB} c_{\beta} \\
m_{+} s_{\alpha} & m^{+-}_{SB} s_{\beta}
\end{array} \right) \ ,
\eqno(A.1)
$$
where $s_{\alpha} = \sin \alpha$, $c_{\alpha} = \cos \alpha$,
and so on.

When we define a rotation
$$
R_{\beta} =  \left(
\begin{array}{ccc}
1 & 0 & 0 \\
0 &  c_{\beta} & -s_{\beta} \\
0 &  s_{\beta} & c_{\beta} \\
\end{array} \right) \ ,
\eqno(A.2)
$$
we obtain
$$
R_\beta^T M =  \left(
\begin{array}{cc}
m^{-+}_{SB} & m_{-} \\
m_{+} \cos (\alpha - \beta) & m^{+-}_{SB} \\
m_{+} \sin (\alpha - \beta) & 0 
\end{array} \right) \ .
\eqno(A.3)
$$
Therefore, the general form (A.1) can always be reduced
into the form with $M_{23} =0$ . Of course, the mixing
angle between $H_{(+)}$ and $\overline{5}_{(+)1}$ in the
model with $M_{23} =0$ is modified by the parameter of 
$\overline{5}_{L(+)} H_{(-)}$. However, it is not essential
in the present model.

On the other hand, it is essential whether the Z$_2$ symmetry
is broken by 
$\overline{H}_{(+)} H_{(-)}$ or $\overline{H}_{(-)} H_{(+)}$.
First, let us see the case where the symmetry is broken only
by $\overline{H}_{(-)} H_{(+)}$ :
$$
M =  \left(
\begin{array}{cc}
m_{SB} & m_{-} \\
m_{+} c_{\alpha} & 0 \\
m_{+} s_{\alpha} & 0
\end{array} \right) \ .
\eqno(A.4)
$$
When we define
$$
R_{\alpha} =  \left(
\begin{array}{ccc}
1 & 0 & 0 \\
0 &  c_{\alpha} & -s_{\alpha} \\
0 &  s_{\alpha} & c_{\alpha} \\
\end{array} \right) \ ,
\eqno(A.5)
$$
we obtain 
$$
R_{\alpha}^T M =  \left(
\begin{array}{cc}
m_{SB} & m_{-} \\
m_{+} &  0  \\
0 & 0
\end{array} \right) \ .
\eqno(A.6)
$$
The mixing matrix $\overline{U}$ among $(\overline{H}_{(-)} \ ,
\overline{H}_{(+)} \ , \overline{5}_{(+)1})$ is given by
$$
\overline{U} = R_{\alpha} R_{\theta} =  \left(
\begin{array}{ccc}
c_{\theta} & s_{\theta} & 0 \\
-s_{\alpha} s_{\theta} &  c_{\alpha} c_{\theta} & -s_{\alpha} \\
-s_{\alpha} s_{\theta} &  s_{\alpha} c_{\theta} & c_{\alpha} \\
\end{array} \right) \ ,
\eqno(A.7)
$$
where 
$$
R_{\theta} =  \left(
\begin{array}{ccc}
c_{\theta} & s_{\theta} & 0 \\
-s_{\theta} & c_{\theta} & 0 \\
0 & 0 & 1 \\
\end{array} \right) \ ,
\eqno(A.8)
$$
with
$$
\tan 2\theta = 
\frac{2m_{SB} m_+}{m^2_{+} -m^2_{-} -m^2_{SB}}
\ ,
\eqno(A.9)
$$
because of 
$$
R^T_{\alpha} M M^T  R_{\alpha}
=  \left(
\begin{array}{ccc}
m^2_{SB} +m^2_{-} & m_{SB} m_{+} & 0 \\
m_{SB} m_{+} & m^2_{+} & 0 \\
0 & 0 & 0 \\
\end{array} \right) \ .
\eqno(A.10)
$$
As we have shown in (2.32)--(2.35), the coefficients 
$\lambda_{1ij}$ of the $R$-parity violating terms 
$\overline{5}_{L1} \overline{5}_{Li} 10_{L(-)j}$ are proportional
to the factor $\kappa = \overline{U}^{(2)}_{13} / \overline{U}^{(2)}_{11}$.
As seen in (A.7) , the case (A.4) leads to $\overline{U}_{13} = 0$, so that
we cannot obtain the effective $R$-parity violating terms 
$\overline{5}_{L} \overline{5}_{L} 10_{L(-)}$.

Of course, although we can obtain the effective $R$-parity violating terms in
the general case with $M_{11} \neq 0$ and $M_{22} \neq 0$, the essential term
to derive the effective $R$-parity violating term is not $\overline{H}_{(-)}
H_{(+)}$, but $\overline{H}_{(+)} H_{(-)}$. In the present paper, in order to
make the essential line of the scenario clear, we have confined ourselves to
investigating only the case with $M_{22} \neq 0$.
Also note that the case without the $\overline{H}_{(-)} H_{(+)}$ term
leads to $\langle \tilde{\nu}_i\rangle =0$ at tree level.

\vspace{7mm}

\centerline{\large\bf Appendix B. Two rank-1 matrix model and
nearly bimaximal mixing} 

\vspace{4mm}
In this Appendix, we investigate the constraint on the two 
rank-1 matrix model with the form

$$
M_\nu =  \left(
\begin{array}{ccc}
f^2_1 & f_1 f_2 & f_1 f_3 \\
f_2 f_1 & f^2_2 & f_2 f_3 \\
f_3 f_1 & f_3 f_2 & f^2_3 \\
\end{array} \right) \ +
\left(
\begin{array}{ccc}
g^2_1 & g_1 g_2 & 0 \\
g_2 g_1 & g^2_2 & 0 \\
0 & 0 & 0 \\
\end{array} \right) \ ,
\eqno(B.1)
$$
which leads to a nearly bimaximal mixing
$$
U_\nu =  \left(
\begin{array}{ccc}
c_{12}\sqrt{1-\varepsilon_{13}^2} & s_{12}\sqrt{1-\varepsilon_{13}^2} 
& \varepsilon_{13} \\
-\frac{s_{12}-c_{12}\varepsilon_{13}}{\sqrt2} &
\frac{c_{12}+s_{12}\varepsilon_{13}}{\sqrt2} &
 -\frac{\sqrt{1-\varepsilon_{13}^2}}{\sqrt2} \\
-\frac{s_{12}+c_{12}\varepsilon_{13}}{\sqrt2} &
\frac{c_{12}-s_{12}\varepsilon_{13}}{\sqrt2} &
\frac{\sqrt{1-\varepsilon_{13}^2}}{\sqrt2} \\
\end{array} \right) \ .
\eqno(B.2)
$$

First, we investigate a general form of $M_\nu$ which gives the neutrino 
mixing (B.2) as follows:
$$
M_\nu = U_\nu D_\nu U^T_\nu \equiv
\left(
\begin{array}{ccc}
a & d_2 & d_3 \\
d_2 & b_2 & c \\
d_3 & c & b_3 \\
\end{array} \right) \ ,
\eqno(B.3)
$$

where $D_\nu \equiv {\rm diag} (m_{\nu_1},m_{\nu_2},m_{\nu_3})$ , and 
$$
a=c^2_{12} (1-\varepsilon_{13}^2) m_{\nu 1} 
+ s^2_{12} (1-\varepsilon_{13}^2) 
m_{\nu 2} + \varepsilon_{13}^2 m_{\nu 3} \ ,
\eqno(B.4)
$$
$$
b_2 = \frac{1}{2}
\left[
(s_{12}-c_{12}\varepsilon_{13})^2 m_{\nu 1} + (c_{12}+s_{12} 
\varepsilon_{13})^2 m_{\nu 2}
+ (1-\varepsilon_{13}^2) m_{\nu 3} \right] \ ,
\eqno(B.5)
$$
$$
b_3 = \frac{1}{2}
\left[
(s_{12}+c_{12}\varepsilon_{13})^2 m_{\nu 1} + (c_{12}-s_{12} 
\varepsilon_{13})^2 m_{\nu 2}
+ (1-\varepsilon_{13}^2) m_{\nu 3}
\right] \ ,
\eqno(B.6)
$$
$$
c = \frac{1}{2}
\left[
(s^2_{12}-c^2_{12}\varepsilon_{13}^2) m_{\nu 1} + (c^2_{12}-s^2_{12} 
\varepsilon_{13}^2) m_{\nu 2}
- (1-\varepsilon_{13}^2) m_{\nu 3}
\right] \ ,
\eqno(B.7)
$$
$$
d_2 = -\frac{1}{\sqrt2}\sqrt{1-\varepsilon_{13}^2}
\left[
c_{12}(s_{12}-c_{12}\varepsilon_{13}) m_{\nu 1} +s_{12}(c_{12}+s_{12} 
\varepsilon_{13}) m_{\nu 2}
+ \varepsilon_{13} m_{\nu 3}
\right] \ ,
\eqno(B.8)
$$
$$
d_3 = -\frac{1}{\sqrt2}\sqrt{1-\varepsilon_{13}^2}
\left[
c_{12}(s_{12}+c_{12}\varepsilon_{13}) m_{\nu 1} +s_{12}(c_{12}-s_{12} 
\varepsilon_{13}) m_{\nu 2}
+ \varepsilon_{13} m_{\nu 3}
\right] \ .
\eqno(B.9)
$$

Next, we rewrite the expression (B.3) into the expression (B.1):
$$
M_\nu =  \left(
\begin{array}{ccc}
\alpha & \delta & d_3 \\
\delta & \beta & c \\
d_3 & c & b_3 \\
\end{array} \right) \ +
\left(
\begin{array}{ccc}
a-\alpha & d_2 - \delta & 0 \\
d_2 -\delta & b_2 - \beta & 0 \\
0 & 0 & 0 \\
\end{array} \right) \ ,
\eqno(B.10)
$$
where $\alpha$, $\beta$ and $\delta$ must satisfy the relations 
$$
\beta = \frac{c^2}{b_3} \ , \ \ \ \alpha = \frac{d^2_3}{b_3} \ , \ \ \  
\delta = \pm \frac{cd_3}{b_3} \ ,
\eqno(B.11)
$$
since the first term of (B.10) is a rank-1 matrix. In order that 
the second term is a rank-1 matrix, the following relation
must be satisfied:
$$
(d_2 - \delta)^2 = (a-\alpha)(b_2 - \beta) \ ,
\eqno(B.12)
$$
which leads to the constraint
$$
0 = a(b_2 b_3 -c^2)-(b_2 d^2_3 + b_3 d^2_2 - 2cd_2 d_3)
= m_{\nu 1}m_{\nu 2}m_{\nu 3} [c^2_{12}-s^2_{12}
(1-2\varepsilon_{13}^2)] \ ,
\eqno(B.13)
$$
for the case $\delta = +cd_3/b_3$ . In the two rank-1 matrix model,
since $m_{\nu 1}m_{\nu 2}m_{\nu 3}=0$, the constraint (B.13) [therefore,
(B.12)] is always satisfied. This means that the two rank-1 matrix 
model (B.1) always has the parameter values which give the nearly bimaximal mixing
(B.2).

\vspace{7mm}
\centerline{\large\bf Appendix C. \ An example of $M_\nu$} 

\vspace{4mm}

We demonstrate an example of the mass matrix 
(4.20).
We take a simple form of the rank-1 matrix $S$ which
satisfies $PSP=P$, 
$$
S =  \left(
\begin{array}{ccc}
1 & 0 & 0 \\
0 & 0 & 0 \\
0 & 0 & 0
\end{array} \right) \ ,
\eqno(C.1)
$$ 
and assume the following form of the neutrino mass matrix $M_\nu$
$$
M_\nu =  m_0 \left(
\begin{array}{ccc}
1 & 0 & 0 \\
0 & 0 & 0 \\
0 & 0 & 0
\end{array} \right)
+ k m_0
\left(
\begin{array}{ccc}
a^2 & a & a \\
a & 1 & 1 \\
a & 1 & 1
\end{array} \right) \ .
\eqno(C.2)
$$
The mass matrix (C.2) gives the maximal mixing
$$
\sin^2 2\theta_{23} \equiv 4 U_{\nu 23}^2 U_{\nu 33}^2=1 \,
\eqno(C.3)
$$
and
$$
U_{\nu 13}=0 \ .
\eqno(C.4)
$$
For $k\simeq 1/2$ and $a^2 \simeq 0$,  the mass matrix (C.2) 
leads to a nearly bimaximal mixing
$$
U_\nu =  \left(
\begin{array}{ccc}
c & s & 0 \\
-\frac{1}{\sqrt{2}}s & \frac{1}{\sqrt{2}}c & -\frac{1}{\sqrt{2}} \\
-\frac{1}{\sqrt{2}}s & \frac{1}{\sqrt{2}}c & \frac{1}{\sqrt{2}} \\
\end{array} \right) \ ,
\eqno(C.5)
$$
where $s=\sin\theta$ and $c=\cos\theta$.
When we put
$$
k=\frac{1}{2}(1+x) \ ,
\eqno(C.6)
$$
we obtain
$$
\begin{array}{l}
 m_{\nu 1}  =  \frac{1}{4} m_0 \left( 4+2x+a^2+a^2 x - 
\sqrt{4 x^2 + 8a^2+ 12 a^2 x +a^4 + 4a^2x^2 + 2a^4 x+a^4 x^2}
\right)   \ , \nonumber \\
 m_{\nu 2}  =  \frac{1}{4} m_0 \left( 4+2x+a^2+a^2 x + 
\sqrt{4 x^2 + 8a^2+ 12 a^2 x +a^4 + 4a^2x^2 + 2a^4 x+a^4 x^2}
\right)   \ , \\
 m_{\nu 3}  =  0 \ , \nonumber
\end{array}
\eqno(C.7)
$$
$$
R \equiv \frac{\Delta m^2_{21}}{\Delta m^2_{32}}
\simeq  - \frac{2(1+\frac{1}{2}x)\sqrt{x^2+2a^2}}{
\left(1+\frac{1}{2}x+\frac{1}{2}\sqrt{x^2+2a^2}\right)}
\ ,
\eqno(C.8)
$$
$$
\tan^2 \theta_{solar} \equiv \frac{U_{\nu 12}^2}{U_{\nu 11}^2}
\simeq \frac{\sqrt{2 a^2+x^2}-x}{\sqrt{2 a^2+x^2}+x}
\ . 
\eqno(C.9)
$$
The values $a/x=1$, $\sqrt{2}$, $\sqrt{3}$ and $2$ give
$\tan^2\theta_{solar}=0.27$, $0.38$, $0.45$ and $0.5$
($\theta_{solar}=27^\circ$, $32^\circ$, $34^\circ$ and
$35^\circ$), respectively.
In order to fit the observed value \cite{solar,kamland,atm}
$$
R_{obs} = 
\frac{6.9 \times 10^{-5} \  {\rm eV}^2}{
2.5 \times 10^{-3} \  {\rm eV}^2}
= 2.76 \times 10^{-2} \ .
\eqno(C.10)
$$
For example, for the case $a/x=2$, by taking $a= 2 x= 0.0092$, 
we  obtain the following numerical results
$$
m_{\nu 1} =0.9954 m_0 \ , \ \ m_{\nu 2}=1.0092 m_0 \ , \ \ m_{\nu 3}=0 \ ,
\eqno(C.11)
$$
$$
U_\nu =\left(
\begin{array}{ccc}
-0.8152 & 0.5791 & 0 \\
0.4095  & 0.5765 & -0.7071 \\
0.4095  & 0.5765 & 0.7071 \\
\end{array} \right) 
\simeq
\left(
\begin{array}{ccc}
-\frac{2}{\sqrt{6}} & \frac{1}{\sqrt{3}} & 0 \\
\frac{1}{\sqrt{6}} & \frac{1}{\sqrt{3}} & -\frac{1}{\sqrt{2}} \\
\frac{1}{\sqrt{6}} & \frac{1}{\sqrt{3}} & \frac{1}{\sqrt{2}} \\
\end{array} \right)  \ ,
\eqno(C.12)
$$
$$
\tan^2 \theta_{solar}= 0.505 \ ,
\eqno(C.13)
$$
together with $R=0.0272$ and $\sin^2 2\theta_{atm}=1$.

A simple example of the mixing matrix $U\equiv U_R^d$ which 
leads to the second term of the expression (C.2) from the
form (C.1) is, for example, given by
$$
U =  \left(
\begin{array}{ccc}
0 & s_\alpha  & c_\alpha \\
-\frac{1}{\sqrt{2}} & -\frac{1}{\sqrt{2}} c_\alpha & 
\frac{1}{\sqrt{2}}s_\alpha \\
\frac{1}{\sqrt{2}} & -\frac{1}{\sqrt{2}} c_\alpha & 
\frac{1}{\sqrt{2}}s_\alpha \\
\end{array} \right)\ ,
\eqno(C.14)
$$
which lead to 
$$
U^T S U =
\left(
\begin{array}{ccc}
0 & 0  & 0 \\
0 &  s_\alpha^2 & s_\alpha c_\alpha \\
0 &  s_\alpha c_\alpha & c_\alpha^2 \\
\end{array} \right)\ ,
\eqno(C.15)
$$
so that
$$
U\cdot P{U^T S U}P\cdot U^T =
U \cdot\left(
\begin{array}{ccc}
0 & 0  & 0 \\
0 &  s_\alpha^2 & 0 \\
0 &  0 & 0 \\
\end{array} \right) \cdot U^T =
\frac{s_\alpha^2}{2} \left(
\begin{array}{ccc}
2 s_\alpha^2 & -\sqrt{2} s_\alpha c_\alpha  & 
-\sqrt{2} s_\alpha c_\alpha \\
-\sqrt{2} s_\alpha c_\alpha &  c_\alpha^2 & c_\alpha^2 \\
-\sqrt{2} s_\alpha c_\alpha &  c_\alpha^2 & c_\alpha^2 \\
\end{array} \right)
\ .
\eqno(C.16)
$$



\end{document}